\documentclass[12pt,preprint]{aastex}

\newcommand{\msun}{$M_{\sun}$}
\newcommand{\lsun}{$L_{B,\sun}$}

\def\bfxi{\mathbf{\xi}}

\shorttitle{Structure of Early-Type Galaxies}

\begin{document}

\title{The Baryon Fractions and Mass-to-Light Ratios of Early-Type Galaxies }

\author{Guangfei Jiang and C.S. Kochanek}
\affil{Department of Astronomy, The Ohio State University, Columbus, OH 43210}


\begin{abstract}
We jointly model 22 early-type gravitational lens galaxies with stellar dynamical measurements 
using standard CDM halo models.  The sample is inhomogeneous in both its mass distributions 
and the evolution of its stellar populations unless the true uncertainties are significantly
larger than the reported measurement errors.  In general, the individual systems cannot constrain 
halo models, in the sense that the data poorly constrains the stellar mass fraction of the halo.  
The ensemble of systems, however, strongly constrains the average stellar mass represented by 
the visible galaxies to $0.026\pm0.006$ of the halo mass if we neglect adiabatic compression, 
rising to $0.056\pm0.011$ of the halo mass if we include adiabatic compression.  Both estimates 
are significantly smaller than the global baryon fraction, corresponding to a star formation efficiency 
for early-type galaxies of $10\%-30\%$.  In the adiabatically compressed models, we find an average
local B-band stellar mass-to-light ratio of $(M/L)_0 = (7.2\pm0.5)(M_{\sun}/L_{\sun})$ that 
evolves by $d\log(M/L)/dz = -0.72\pm0.08$ per unit redshift. Adjusting the isotropy of the 
stellar orbits has little effect on the results.  The adiabatically compressed models are
strongly favored if we impose either local estimates of the mass-to-light ratios of early-type 
galaxies or the weak lensing measurements for the lens galaxies on 100~kpc scales as
model constraints. 
\end{abstract}


\section{Introduction}

The presence of dark matter in early type galaxies is clear on large scales, based on
both weak lensing (e.g. Kleinheinrich et al. 2006, Mandelbaum et al. 2006) and X-ray (e.g. Humphrey et al. 2006)
studies. The distribution of the dark matter and the mass fraction represented by the stars
are less well-determined because of the difficulties in measuring early-type galaxy structure in the transition
region between the stars and the dark matter.  Stellar kinematic studies of the central regions,
when compared to estimates of stellar mass-to-light ratios, have argued either that there is
little dark matter inside an effective radius (e.g. Gerhard et al. 2001) or that there
is a substantial dark matter fraction (e.g. Padmanabhan et al. 2004).  The significance
of these differences depends on the reliability of estimating the stellar mass from
combinations of photometry, spectroscopy and population synthesis models.
Studies on larger scales using planetary nebulae,  have found examples of galaxies with 
falling rotation curves (Romanowsky et al. 2003), while the globular clusters in one
of these systems show a flat rotation curve (Pierce et al. 2006).  Surveys of structure
with gravitational lenses (e.g. Rusin \& Kochanek 2005, Treu et al. 2006) indicate that
the typical lens has a flat rotation curve on scales of 1--2$R_e$, but the interpretation
of the scatter around the mean structure has been used to argue for both inhomogeneity
(e.g. Treu \& Koopmans 2002a, Kochanek et al. 2006) and homogeneity  (e.g. Rusin \& Kochanek 2005,
Koopmans et al. 2006) in the mass distributions.

In this paper we reanalyze a sample of 15 lenses from the Sloan Lens ACS Survey (SLACS,
Bolton et al. 2006) and 7 lenses from the Lens Structure and Dynamics Survey (LSD, 
Koopmans \& Treu 2003) that have both mass estimates from the lens geometry and
central velocity dispersion measurements.  Koopmans et al. (2006) analyzed the
sample using a simple, global power law model, $\rho \propto r^{-\gamma}$, for
the mass distribution to find a mean slope of $\gamma=2.01^{+0.02}_{-0.03}$
where $\gamma=2$ corresponds to a flat rotation curve.  With a nominal 
scatter in the slope of only $0.07$, Koopmans et al. (2006) argue that the
halo structures appear to be fairly homogeneous.  It is difficult, however,
to relate these power law models to theoretical models of halos or to evaluate
the significance of the scatter in the slope.  Additionally, the models
are somewhat unphysical because they allow mass distributions
more compact that the luminous galaxy.

Here we reanalyze the SLACS and LSD lens samples using a more physical mass model that combines a \citet{hern90} profile for the stars with a 
Navarro, Frenk \& White (1996, NFW) model for the dark matter. By comparing the mass inside the Einstein ring of the galaxies ($M_E(<R_E)$)
 with the mass needed to produce the observed velocity dispersion, we can estimate the stellar mass fraction and the
 stellar mass-to-light ratio explicitly. We use a Bayesian formalism that allows us to quantitatively address the homogeneity of the sample.
We review the data and describe our mass models and analysis techniques in \S\ref{sec:method}. In \S\ref{sec:result}, we discuss the models of the 
individual systems (\S\ref{sec:indiv}), the homogeneity of the sample (\S\ref{sec:hom}), and finally, the stellar mass fraction, 
the  mean stellar mass-to-light ratio and the evolution of the mass-to-light ratios (\S\ref{sec:barma}). We summarize our results in \S\ref{sec:dis}.

\section{Data and Method}
\label{sec:method}
\subsection{Data}
\label{sec:data}
In this paper we reanalyze the data for 15 lenses from the Sloan Lenses ACS Survey (SLACS) 
and 7 lenses from the Lens Structure and Dynamics Survey (LSD).
We neglect two lenses, Q0957+561 and Q2237+030,  from the LSD,
since Q0957+561 is in a cluster and Q2237+030 is a barred spiral galaxy, 
to leave us with a sample of 22 galaxies with measured 
velocity dispersions, effective radii, Einstein radii, enclosed masses and rest frame B band magnitudes taken from the original analyses.
For convenience, we summarize the data in  Table \ref{tab:tab1}, particularly since the equivalent table in \citet{tkbbm06} 
contains ordering errors. For consistency we have adjusted all the data to a flat $\Lambda$CDM cosmological
model with $\Omega_m=0.3$, $\Omega_{\Lambda}=0.7$ and $\rm{H_0=70~km~s^{-1}~Myr^{-1}}$. 

\subsection{Method of Analysis}
\label{sec:anal}
We model the lenses with the two component mass model for lenses introduced by \citet{keeton01},
which was also used for early-type galaxies in the SDSS by \citet{padman04}. 
It consists of a \citet{hern90} model for the luminous galaxy  and a Navarro, Frenk \& White (1996, NFW) profile for the dark matter halo.
 By using this physically motivated model rather than the simple power law normally used by the LSD/SLACS studies, we both better connect the results
to theoretical halo models, and avoid models in which the dark matter  can be more centrally concentrated than the stars.
In essence, we use the mass enclosed by the Einstein radius of the lens to set the virial mass $M_{vir}$, the stellar velocity dispersion to 
determine the stellar mass fraction $f_*$, and theoretical halo models to constrain the halo concentration $c$. Finally, 
by comparing the derived stellar mass to the observed luminosity, we can estimate the rest frame B-band mass-to-light ratio ${M_{\ast}/ L}$ 
and its evolution.

The \citet{hern90} model used for the luminous lens galaxy is defined by
\begin{equation}
\rho_H(r) = {M_{\ast}\over{2\pi}} {r_H\over{r(r+r_H)^3}},
\end{equation}
where the scale length $r_H = 0.551R_e$ is matched to the measured de Vaucouleurs profile 
effective radius $R_e$ and the stellar mass $M_{\ast} = f_* M_{vir}$ is related to the virial mass by the cold baryon/stellar mass fraction
$f_*$.
The NFW profile \citep{nfw96} used to model the initial dark matter halo is defined by
\begin{equation}
\rho_N(r) = {M_{dm}\over{4\pi f(c)}}~{1\over{r(r+r_s)^3}},
\end{equation}
where the scale length $r_s$ is related to the virial radius by $c = r_{vir}/r_s$,
$f(c) = \ln(1+c) - c/(1+c)$, and
$M_{dm} = (1-f_*)M_{vir}$ is the mass in dark matter.
The average concentration was modeled by
\begin{equation}
\label{eqn:c1}
c = {9\over 1+z}~\left({M_{vir}\over 8.12\times 10^{12}~h M_{\sun}  } \right)^{-0.14},
\end{equation}
and the individual halos have a log-normal dispersion in their concentrations of 
$\sigma_c = 0.18$ (base 10) around the average \citep{bullock01}.
These initial models neglect the compression of the dark matter density profile by the more concentrated baryons.
We estimated the changes in  the dark matter 
distribution using the adiabatic compression model of \citet{blumenthal86}. This approximation may exaggerate the compression 
\citep{gnedin04}, so we should regard our compressed and uncompressed results as bounding the possible effects of adiabatic compression.

The observations provide two constraints, 
the mass inside the Einstein radius, and the stellar velocity dispersion. For any value of $c$ and $f_*$, we use
the projected mass inside the Einstein radius to determine $M_{vir}$ (which also determines $r_{vir}$),
then use the spherical Jeans equation and a constant orbital isotropy $\beta$ to compute the velocity dispersion expected for the
measurement aperture. The effects of seeing were modeled using a Gaussian PSF with the observed FWHM of the observations.
 Given the estimated dispersion
$\sigma_{i,model}$, the measured dispersion $\sigma_i$ and its uncertainties $e_{\sigma i}$ for galaxy $i$, we estimate a goodness of fit
${\chi_i}^2(\sigma_i)=(\sigma_{i,model}-\sigma_i)^2/e_{\sigma i}^2$.
 We model the mass-to-light ratios of the stars using the standard power law (e.g. \citealt{vanf96,treu01,rk05,ktbbm06}),
\begin{equation}
\log\left({M_{\ast} \over L}\right) = \log\left({M_{\ast} \over L}\right)_0 + z\left({d \log(M_{\ast}/L)\over d z}\right)
\end{equation}
where $({M_{\ast} / L})_0$ is the value today and ${d \log(M_{\ast}/L)/ d z}$ is the rate at which it changes with redshift $z$. 
This in turn defines a goodness of fit $\chi_i^2((M/L)_i)$ 
with which the model fits the logarithm of the mass-to-light ratio $(M/L)_i$ of galaxy $i$,
defined by the ratio of the estimated stellar mass (a model parameter) to the observed luminosity,
given its uncertainties $e_{Li}=\Delta(\log(M/L)_i= \Delta L_i/(\ln(10)L_i)$.  
These two terms define a probability of fitting the velocity
dispersion $P(\sigma_i|\bfxi) = \exp(-\chi_i^2(\sigma_i)/2)/\sqrt{2\pi}e_{\sigma i}$ and the mass-to-light ratio 
$P((M/L)_i|\bfxi) = \exp(-\chi_i^2((M/L)_i)/2)/\sqrt{2\pi}e_{L i}$ given the model parameters $f_*$, $c$,
$(M_*/L)_0$ and $d\log(M/L)/dz$ which we abbreviate as $\bfxi$.  Combining the two terms, we have
the probability of the model fitting the data $D_i$ for galaxy $i$
\begin{equation}
\label{eqn:eqn2}
 P_i\left(D_i\big|{\bf\xi}\right) = P\left(\sigma_i|{\bf \xi}\right) P\left((M/L)_i| {\bf \xi}\right).
\end{equation}

In addition to the measurement errors listed in Table \ref{tab:tab1}, we should also consider sources of systematic 
errors. The essence of the method is to compare the mass inside the Einstein ring $M(<R_e)$ to a virial mass 
estimate from the velocity dispersion ${\sigma}_v^2R/G$.  We can identify five sources of systematic errors. 
First, while there is little uncertainty in $M_E$, some of the mass may be projected 
surface density from either a parent group halo to which the lens belongs, or from another along the line of sight. 
The extra density, $\kappa=\Sigma/\Sigma_c$ in dimensionless units, modifies the mass inside the Einstein radius by 
$\pi\kappa R_E^2 \Sigma_c$, so we can think of its effects as a systematic error in interpreting $\sigma_v$ of 
$\rm e_\sigma=\sigma_{\kappa}/2$.  The full probability distribution of $\kappa$ is skewed to positive 
values (e.g. \citealt{th03}), but we will ignore this problem and assume $\sigma_{\kappa}\backsimeq 0.05$ since the
positive tail of the distribution is associated with detectable objects (galaxies and clusters).  
This systematic error also affects estimates of the mass-to-light ratios. Second, 
there are $1-10\%$ uncertainties in the galaxy effective radius measurements which contribute uncertainties of 
$0.5\%$ to $5\%$  to our interpretation of the velocity dispersion.  Third, the measured velocity dispersion 
is a Gaussian fit to the spectrum, which is not identical to the rms velocity appearing in the Jeans equation 
(e.g. \citealt{bs87}).  The difference can be estimated from the typical Gaussian-Hermite coefficients 
$|h_4|\backsimeq0.02$ \citep{bsg94} as a fractional error in $\sigma_v$ of order $\sqrt{6}|h_4|\backsimeq0.05$ 
in the velocity dispersion (e.g. \citealt{vf03}).  Fourth, non-sphericity, (somewhat to our surprise) leads 
to negligible systematic errors provided we use the intermediate scale length (the geometric mean of the 
semi-major and minor axes), at least in the limit of the tensor virial theorem. It leads to large errors 
if any other scale length is used.  \citet{Barnabe07} have taken the first steps towards 
removing these two dynamical problems, although they are restricted to oblate two-integral models
which may not be appropriate for massive elliptical galaxies.  
Finally, calibration errors in the velocity dispersions contribute 
fractional errors of order 0.03 (see \citealt{bernardi03a}).  Combining all these contributions in 
quadrature, which corresponds to assuming a Gaussian model for each systematic error, we estimate that the typical
systematic uncertainty to interpreting the velocity dispersions is approximately $8\%$ with the exact value 
depending on the uncertainties in the effective radius.

Our statistical methods are chosen so that we can understand the homogeneity of the lens galaxies 
in either their evolution or their dynamical properties and estimate their average properties in 
the presence of inhomogeneities.  We will analyze the results using two Bayesian methods.  In the
first method, we will fit the data while simultaneously estimating the systematic errors in the
velocity dispersion and the mass-to-light ratio.  When combined with the measurement errors, 
these define new uncertainty estimates for the data which we will call the ``bad'' case errors
in comparison to the original uncertainties (the ``good'' case).  These broadened uncertainties
can be representative of either true systematic uncertainties, such as the ones we discussed above
for the dynamical measurements, or indicative of inhomogeneities in the structure or evolution of
the galaxies. In the second method we
will compare these two cases using the approach outlined in \citet{bo97} to determine the 
degree to which the sample homogeneous or heterogeneous.  In this method, 
we assume that there are probabilities $p_\sigma$ and $p_L$ that the galaxies have 
homogeneous structures or evolutionary histories in the sense that the scatter in the measurements
is simply determined by the ``good'' measurement errors.  There are then probabilities $1-p_\sigma$ and $1-p_L$
that the galaxies are not a homogeneous group in either their structure or their evolution, where
we characterize this by assuming that the uncertainties in the velocity dispersion and the mass-to-light 
ratio are significantly broadened to be the ``bad'' measurement errors. In essence, we are determining
the relative probabilities of the stated measurement errors and our estimate of the 
true uncertainties from the first method.  Both approaches provide uncertainties on the average
properties of the sample that account for potential inhomogeneities, although the second method
is a better formal approach since it can reject individual objects. 

In the first approach, we will estimate the fractional systematic errors $e_\sigma$ and
$e_L$ in the velocity dispersion and luminosity.  The $\chi^2$ expressions are modified to use uncertainties
of $e_{\sigma i} \rightarrow \sqrt{e_{\sigma i}^2+e_\sigma^2\sigma_i^2}$
and $e_{L i} \rightarrow \sqrt{e_{Li}^2+e_L^2}$ for the velocity dispersions and the logarithm of
the mass-to-light ratios respectively.
We assume logarithmic priors for $f_*$, $\left({M_{\ast}/ L}\right)_0$, and
$\left({d\left(M/L\right)/dz}\right)$, and the theoretical prior defined by Eqn.~(\ref{eqn:c1}) for the
concentration $c_i$.  Note that we are forcing all galaxies to have the same concentration, which
should have no significant impact given the scales we are studying.
The priors for the systematic errors, $P(e_\sigma)=1/\sqrt{\langle e_{\sigma i}^2 \rangle + e_\sigma^2\sigma_i^2}$
and $P(e_L)=1/\sqrt{\langle e_{Li}^2\rangle + e_L^2}$, naturally switch between uniform priors
for systematic errors small compared to the mean square measurement errors ($\langle e_{\sigma i}^2 \rangle$
and $\langle e_{Li}^2 \rangle$) and logarithmic priors for large systematic errors.
The resulting probability distribution for the fractional errors is then
\begin{equation}
  P(e_\sigma, e_L |D) \propto P(e_\sigma)P(e_L)
    \int d\mathbf{\xi} P(\mathbf{\xi}) {\prod_{i}}
      P\left(D_i|e_\sigma,\mathbf{\xi}\right)
       P\left(D_i|e_L,\mathbf{\xi}\right)
\label{eqn:e1e2}
\end{equation}
where $P\left(D_i|e_\sigma,\mathbf{\xi}\right)$ and $P\left(D_i|e_L,\mathbf{\xi}\right)$ are
the probability distributions modified by the addition of the systematic errors $e_\sigma $ and
$e_L$.  We then use these systematic error estimates to define the uncertainties used for
the ``bad'' case in our second formalism.

The second, \citet{bo97} approach properly weights all combinatoric possibilities of the individual systems 
being members of a homogeneous sample or not.  Let $P_{Gi}(\sigma_i|\bfxi)$ and $P_{Gi}((M/L)_i|\bfxi)$
be the probabilities of the data given the parameters for galaxy $i$ if it is a member of a homogeneous 
group based on the measured, ``good'' uncertainties, and $P_{Bi}(\sigma_i|\bfxi)$ and $P_{Bi}((M/L)_i|\bfxi)$ 
be the probabilities if it is not and we should be using the ``bad'' uncertainties based on the systematic
error estimates derived from our first method.  The \citet{bo97} provides estimates of the relative
likelihoods describing either the full sample or the individual systems by the 
``good'' or ``bad'' data model.
If we want the Bayesian probability distribution for the parameters $\bfxi$ properly  weighted
over all possible group membership combinations, we find that
\begin{equation}
P(\mathbf{\bfxi}|D) \propto P(\mathbf{\bfxi}) \int dp_\sigma dp_L {\prod_{i}} F_i
\label{eqn:p1}
\end{equation}
where
\begin{equation}
    F_i = \left[ p_\sigma P_{Gi}(\sigma_i|\bfxi) + (1-p_\sigma) P_{Bi}(\sigma_i|\bfxi)\right]
          \left[ p_L P_{Gi}((M/L)_i|\bfxi) + (1-p_L) P_{Bi}((M/L)_i|\bfxi)\right]
\end{equation}
and where $P(\mathbf{\xi})$ sets the prior probability distributions for the parameters. We assume a uniform priors 
for $p_\sigma$ and $p_L$.  We obtain the probability distribution for any parameter by marginalizing  Eqn.~(\ref{eqn:p1}) 
over all other variables and then normalizing the total probability to unity.  We can also
estimate the probability that the sample is homogeneous in either its structural or evolutionary
properties as
\begin{equation}
   P(p_\sigma,p_L| D_i) \propto \int d\bfxi P(\bfxi) \Pi_i F_i
  \label{eqn:egb}
\end{equation}
and the probability that a particular galaxy is in the dynamically homogeneous class is
\begin{equation}
\label{eqn:aob}
P(\sigma_i \in \hbox{homogeneous} |D) = {A_i \over A_i+B_i}
\end{equation}
where 
\begin{eqnarray}
\label{eqn:ab}
  A_i =\int d\mathbf{\xi}P(\mathbf{\xi})\int dp_\sigma dp_L  p_\sigma P_{Gi}(\sigma_i) 
              \left[ p_L P_{Gi}((M/L)_i|\bfxi) + (1-p_L) P_{Bi}((M/L)_i|\bfxi)\right] \Pi_{i\ne j} F_j \quad\hbox{and}\\
  B_i =\int d\mathbf{\xi}P(\mathbf{\xi})\int dp_\sigma dp_L  (1-p_\sigma) P_{Bi}(\sigma_i) 
              \left[ p_L P_{Gi}((M/L)_i|\bfxi) + (1-p_L) P_{Bi}((M/L)_i|\bfxi)\right] \Pi_{i\ne j} F_j. \nonumber
\end{eqnarray}
A similar set of expressions gives the probability that the galaxy is in the set of galaxies with
a homogeneous evolutionary history.

\section{Results}
\label{sec:result}

We divide our discussion of the results into three subsections. First, we present the results for the 
individual galaxies. Next, we discuss the homogeneity of the structural properties of the galaxies. 
Finally, we estimate the stellar mass fraction, mass-to-light ratios and the rate of galaxy evolution.

\subsection{Properties of Individual Galaxies}
\label{sec:indiv}

Figure ~\ref{fig:figure1} shows contours for the goodness of fit of the models to 
the velocity dispersion, measured for each galaxy as a function of the stellar mass fraction $f_*$ and the 
concentration $c$ once we have normalized the mass inside the Einstein radius. For these calculations, we 
have included our estimates of the systematic errors in the velocity dispersions but used the stated 
uncertainties in the luminosities.
Note that the dispersion measurements cannot determine the halo 
concentrations but the goodness of fit contours always pass through the region set by our prior on the 
concentration.  The permitted stellar mass fractions vary widely between objects. Three of the 22 objects, 
SDSS J$0737+321$, SDSS J$1250+052$ and PG$1115+080$, appear to require mass distributions that are more 
centrally concentrated than the stars, in the sense that the best fits for $f_* \leq 1$ have $\chi^2>2$. 
This is also seen in the LSD models for PG$1115+080$ \citep{tk02a}, where the only models consistent 
with both the lensing constraint and the estimated velocity dispersion are more centrally concentrated 
than the stars.  A fourth lens, SDSS J1627$-$005, is only marginally consistent with $f_* \leqslant 1$ .
Of the remaining 18 galaxies, eleven are consistent with $f_* = 1$ ($\Delta\chi^2<1$), and seven are not. 
Four of these eleven galaxies have enormous parameter uncertainties.  One problem for many SLACS lenses is that 
the scales of the velocity dispersion aperture/effective radius differ little from the observed Einstein 
radius, which limits the leverage for constraining the mass profile. 

Figure~\ref{fig:figure3} shows the goodness of fit to the mass-to-light ratio of each galaxy given the
best fit model for the average evolution of the sample.  Most of the galaxies are consistent with this 
best fit model for the mass-to-light ratio and its 
evolution (\S\ref{sec:barma}).  The mass-to-light ratios of the sample appear to be more uniform than the 
dynamical properties, probably for the same reasons that there is little scatter in the fundamental plane 
(see  \citealt{bernardi03b}).  However, there are three 3$\sigma$ outliers in the sample, SDSS J1420+602, 
SDSS J1250+052 and H1543+535, all of which have very low M/L ratios compared to the other galaxies.
Note that only one of these, SDSS J1250+052, is also an outlier in the dynamical fits. This is not unique 
to our approach, since our mass-to-light ratio for H1543+535 is comparable to that in \citet{tk04}. In 
Figure 6 of \citet{tkbbm06}, they also find significantly lower mass-to-light ratios for SDSS J1420+602 
and SDSS J1250+052  than they do for the other SLACS members. One possible solution is that the lens masses 
are significantly mis-estimated due to contamination from a group or cluster halo, but only H1543+535 has 
a neighboring, bright galaxy and it is sufficiently distant to only modestly perturb the estimated mass. 

\subsection{Homogeneity}
\label{sec:hom}

The broad uncertainties and occasional outliers mean that it is important to have 
a quantitative approach to determining the homogeneity of the sample and to appropriately 
weight each object when determining mean 
properties. This is why we introduced the Bayesian frameworks of \S\ref{sec:method}. 
Fig.~\ref{fig:figure6} shows our estimates of the fractional systematic errors from our
first analysis method (Eqn.~\ref{eqn:e1e2}).  The best fit estimates for the fractional
systematic errors in the velocity dispersion and luminosity are $e_\sigma \simeq 0.1$
and $e_L\simeq 0.18$.  The reported measurement errors lie well outside the 99.7\%
likelihood region.   For the dynamical errors, the best fit systematic errors are 
quite consistent with our prior estimates based on simple considerations about the
dynamical data. 

Fig.~\ref{fig:figure5} shows that the results for the homogeneity of the sample are very sensitive to the assumed 
uncertainties.  If we simply used the stated measurement errors, then the probability that the sample is
homogeneous in its dynamical properties 
(i.e. that the ``good'' uncertainty estimates are correct and the scatter is due only to measurement error) is 
$p_\sigma \le{24\%}$ and that it is homogeneous in mass-to-light ratio
evolution is $p_L \le{14\%}$.  Many objects have low probabilities of belonging to either a 
homogeneous dynamical subset (SDSS J1627--0053 with $p_\sigma = 0.005$, SDSS J1250+0523 with $p_\sigma = 0.010$, 
SDSS J0737+321 with $p_\sigma = 0.010$, PG1115+080 with $p_\sigma = 0.029$, and SDSS J1420+602 with $p_\sigma = 0.042$)
or a homogeneous evolutionary subset 
(SDSS J1250+0523 $p_L \approx 0$, H1543+535 $p_L \approx 0$, SDSS J0912+002 $p_L = 0.001$, SDSS J1420+602 $p_L = 0.001$ and 
MG1549+305 $p_L = 0.001$).
Not surprisingly, these objects are also
outliers in the individual fits from the previous section.  If we include our estimates of the
systematic errors in interpreting the dynamical measurements, then the probability that the sample
is homogeneous in its dynamical properties rises to $p_\sigma \ge 40\%$, but the probability of
a homogeneous evolutionary population remains small at $p_L\le{14\%}$. 
With the inclusion of the systematic error estimates, the objects with the lowest probabilities of belonging 
to the homogeneous dynamical subset are PG1115+080 (with $p_\sigma = 0.48$), 
SDSS J1250+0523 (with $p_\sigma = 0.49$), SDSS J1627--0053 (with $p_\sigma = 0.52$),
SDSS J0737+321 (with $p_\sigma = 0.52$), and H1417+526 (with $p_\sigma = 0.55$).  
These estimates strongly indicate that either the SLACS/LDS lens populations are inhomogeneous or
that the measurement errors underestimate the true uncertainties.  In a few cases, these problematic
lenses show some evidence for disks (SDSS J1420+602, MG1549+305).  
 
In sum, the SLACS/LDS galaxies are homogeneous in neither their dynamical nor their evolutionary
properties if we take the measurement errors at face value.  It is likely that most of the
problem for the dynamical measurements is that the systematic errors in interpreting velocity
dispersions are significant and need to be included in any analysis
of the dynamics of lenses.  One of these systematic errors, surface density contributions from
structures other than the lens galaxy, also produces systematic errors in the mass-to-light
ratio, with $\sigma_M = \sigma_\kappa \simeq 0.05$, but this is much too small to explain 
the spread in the mass-to-light ratios.  This problem is probably caused by a combination of underestimated uncertainties
in the luminosities and true variance in the evolutionary history of early-type galaxies.
Rusin \& Kochanek (2005) and \citet{tkbbm06} had found earlier that the lens sample was better fit by allowing
a range for the mean redshift at which the stars formed than by assuming a single value,
and in this analysis a range of formation redshifts would lead to non-zero systematic errors 
in the mass-to-light ratio.

\subsection{The Stellar Mass Fraction and Mass-to-Light Ratio}
\label{sec:barma}

We can combine the galaxies to make joint estimates of the stellar mass fraction $f_*$, the mass-to-light 
ratio $M_{\ast}/L$ at $z=0$ and its evolution.  We considered both Bayesian frameworks so that
either the uncertainties are broadened to make the results consistent with all the lenses 
(method 1, Eqn.~\ref{eqn:e1e2}) or  the 
outliers in the sample are properly down weighted in the analysis (method 2, Eqn.~\ref{eqn:p1}) 
The two methods give similar results, so we only present the detailed results from the second
Bayesian method.
Fig.~\ref{fig:figure7}, shows the estimated stellar mass fraction $f_*$ for both the individual galaxies 
and the sample as a whole, and for orbital anisotropies of $\beta=-1/3, 0$ and 1/3, where 
$\beta=1-{{\sigma_t}^2/{\sigma_r}^2}$ is related to the ratio of the tangential $\sigma_t$ 
and radial $\sigma_r$ velocity dispersions.  For isotropic, adiabatically compressed models, we find 
$f_*=0.056\pm0.011$, and like \citet{ktbbm06}, we find that that the isotropy has little effect on 
the inferred mass distribution. The stellar mass fraction is significantly
lower than the global baryon fraction of $0.176^{+0.006}_{-0.019}$ from the WMAP CMB anisotropy 
measurements \citep{spergel06}.  If we do not include the adiabatic compression of the halo,  
the stellar mass fraction drops to $0.026\pm0.006$, again with little dependence on the isotropy $\beta$. 
While the uncompressed models have less dark matter in the central regions, the total halo mass
is much larger than in the compressed models.
To the extent that adiabatic compression occurs, but the \citet{blumenthal86} model exaggerates its degree 
\citep{gnedin04}, reality is intermediate to these two extremes.

We also fit the mass-to-light ratio as $\log(M_{\ast}/L)=\log a+bz$, where $a=(M_{\ast}/L)_{0}$ is the 
mass-to-light ratio at $z=0$ and $b=d(\log(M_{\ast}/L))/dz$ is its evolution with redshift.
Fig.~\ref{fig:figure8}, shows the likelihood contours for these two parameters for both compressed and 
uncompressed models.  For the compressed isotropic models, we find $a=(7.2 \pm 0.5) M_{\sun}/L_{\sun}$ 
and $b=-0.72\pm 0.08$. This agrees with the local value of $a = (7.3\pm 2.1) M_{\sun}/L_{\sun}$
from \citet{gerhard01} that was used by \citet{tkbbm06}. It also agrees with the \citet{tkbbm06} estimate 
for the rate of the evolution $b=-0.69\pm 0.08$. 
Changing the isotropy over the range $\beta=-1/3, 0, 1/3$ has little effect, while the model without 
adiabatic compression requires higher normalizations for the mass-to-light ratio ($10.0\pm0.3$) and 
slightly slower rates of evolution.  Our analysis includes 2 lenses (PG1115+080 and H1543+535) that 
were not used by \cite{tkbbm06}, but excluding them from the analysis has little effect on the mass-to-light 
ratios. There are no significant changes in $(M_{\ast}/L)_0$ and $d\log(M_{\ast}/L)/dz$ if we neglect these two lenses.

\section{Discussion}
\label{sec:dis}

We reanalyzed the data from the SLACS and LSD surveys of gravitational lenses with velocity dispersion 
measurements.  Our mass distribution consists of a Hernquist model for the luminous galaxy embedded in a 
theoretically constrained NFW halo model.  We investigated the homogeneity of the sample, the stellar mass 
fraction $f_*$, the local ($z=0$) stellar mass-to-light ratio  $(M_{\ast}/L)_0$ and its evolution
$d(\log(M_{\ast}/L))/dz$.  As in the earlier study by \citet{ktbbm06}, we found that the effects of orbital 
anisotropy on both the stellar mass fraction and the mass-to-light ratio are small.

In most cases, a central velocity dispersion measurement provides only weak a constraint on halo structure in 
the physically interesting region. Typical limits on the mass fraction represented by the stars have logarithmic 
errors of order 0.5 dex. While this appears to contradict the conclusions of (for example) \citet{ktbbm06}, this is not the 
case. \citet{ktbbm06} fit mass models where $\rho\varpropto r^{-\gamma}$, and find values in the range 
$1.8<\gamma<2.3$.  Fig.~\ref{fig:figure9} shows the expected range of this slope for our models of 
SDSS~J0037--0942, where we estimated the slope by fitting the $\rho\varpropto r^{-\gamma}$ power law  model 
to the projected mass distribution over a radial baseline of $R_e/8$ to $R_E$ that approximates 
the leverage in using stellar dynamics combined with gravitational lensing to determine halo 
structure.  For this typical lens, the variations in $\gamma$ of $1.6 \la \gamma \la 2.06$ are comparable 
to the system-to-system spread in $\gamma$ observed for the SLACS systems (\citet{ktbbm06}).  
Thus, the spread in $\gamma$ observed for individual SLACS/LSD lenses is comparable to the range
of values found in our halo models, so strong conclusions about 
halo structure from these system will depend on averages over the samples rather than the results for 
individual lenses. 

The critical issue for determining the sample average properties is the degree to which the populations 
are homogeneous. A heterogeneous sample cannot easily be averaged to determine mean properties. We 
find the probability of homogeneity is very sensitive to the uncertainties in both the velocity
dispersion and the luminosity. If take the measurement errors at face value, there is a low
probability of homogeneity in either dynamical structure ($p_\sigma\le{20\%}$) or evolutionary history
($p_L\le{15\%}$).  Many lenses such as SDSS J1250+052, H1543+535, SDSS J1420+601, SDSS J0912+002, 
MG2016+112, and MG1549+305 have low ($<10\%$) likelihoods
of belonging to a homogeneous sample.  The primary problem is probably that there are significant systematic 
uncertainties that must be included with the measurement errors.  Simple considerations show that
typical systematic errors in interpreting the velocity dispersions should be large compared to the
measurement errors, 8\% versus 5\%, and adding these estimated systematic uncertainties greatly increases the likelihood of
dynamical homogeneity.  Sources of systematic error in the mass-to-light ratio are less amenable
to simple arguments, but should certainly include the dispersion in the average star formation epoch
of early-type galaxies found in an earlier analyses of galaxy evolution with lenses by Rusin \& Kochanek
(2005) and \citet{tkbbm06}.  If we simply analyze the data to determine the most likely systematic errors, we find that
we must include fractional systematic errors of approximately 10\% in the velocity dispersion estimates 
and 19\% in the mass-to-light ratio estimates in order to make the sample consistent with the 
hypothesis of homogeneity.

Once we account for the inhomogeneity or systematic errors in the sample, we can evaluate sample averages 
that properly account for these problems.  We find that the halo mass fraction represented by the baryons in 
stars is $f_* = 0.056\pm0.011$ if we adiabatically compress the dark matter and $f_* = 0.026\pm0.006$ if we 
do not.  These results are comparable to similar the range of estimates that relied 
on stellar population models to estimate the stellar mass.  For example, \citet{lfl06} obtained a stellar mass 
fraction of $\backsim 8\%$ by fitting monolithic collapse models to 2000 SDSS galaxies, \citet{hhylg05} found 
$f_* = 0.065^{+0.010}_{-0.008}$ using weak lensing, and Mandelbaum et al. (2006) found $f_*=0.03^{+0.02}_{-0.01}$\% 
using weak lensing.  The results in these studies depend on the assumed IMF --  the \citet{hhylg05} estimate 
drops to $f_* = 0.035^{+0.005}_{-0.004}$ if the initial mass fraction of the stars is changed from a standard 
Salpeter IMF to a scaled Salpeter IMF.  Our results probably bound the stellar mass fraction since the 
Blumenthal et al. (1986) model we used may overestimate the amount of adiabatic compression (Gnedin et al. 2004).  
In all our models, the stellar mass fraction is much smaller than the cosmological baryon mass fraction 
$\Omega_b/\Omega_m=0.176^{+0.006}_{-0.019}$ from WMAP \citep{spergel06}, which means that the star 
formation efficiency ($f_* \Omega_m/\Omega_b$) of early-type galaxies is only 15--30\%.   The remaining 
baryons must remain as gas distributed on the scale of the halo or its parent (group) halo. This discrepancy 
appears to be a common problem for any baryon accounting for normal galaxies (e.g. \citealt{fukugita04}) and a 
significant constraint on star formation efficiency.

Analysis of the evolution of early-type galaxies with redshift, whether using the fundamental plane 
(e.g. \citealt{jorgen99,franx00,tkbbm06}), dynamical mass estimates (e.g. \citealt{md06}), or gravitational 
lens data alone (e.g. \citealt{rk05}), have consistently observed a steady brightening of early-type galaxies 
with look-back time, albeit with modest disagreements as to the rate. Here we use a hybrid method, fitting 
the stellar mass-to-light ratios inferred from mass models of gravitational lenses with
stellar dynamical data, to find that $(M_{\ast}/L)_0 = (7.2 \pm 0.5) M_{\sun}/L_{\sun}$ and 
$d(\log(M_{\ast}/L))/dz = -0.72\pm 0.08$ for the compressed models. The mass-to-light ratio is comparable to the 
local value of $(M_{\ast}/L)_0 = (7.3\pm 2.1) M_{\sun}/L_{\sun}$ from \citet{gerhard01} which was
adopted by \citet{tkbbm06} in their analysis of this data. The mass-to-light ratio evolution rate is also 
close to the value $d\log(M_{\ast}/L)/dz=0.69\pm0.08$ from \citet{tkbbm06},  and marginally larger than
the estimates by \citet{rk05} and \citet{md06}. As pointed out by \citet{rk05}, the differences in
evolution rates are partly due to different approaches to weighting the contribution of each lens
to the analysis, but at least our Bayesian approach 
carries out these weightings in an objective fashion.  

In our basic analysis we cannot distinguish between the adiabatically compressed
and uncompressed models.  In essence, we can obtain the same mass distribution either using a high
stellar mass-to-light ratio and a more extended halo or the reverse.  If we impose the locally estimated
mass-to-light ratio as a constraint, then the adiabatically compressed model is favored (6 to 1). 
 We can also use mass measurements on much larger scales to distinguish
the two models because the total halo mass is larger in the uncompressed model.  In particular,
we can calculate the weak lensing $\Delta\Sigma$ and compare it to the measurement by Gavazzi
et al. (2007, also see Mandelbaum et al. 2006) for an overlapping sample of SLACS lenses where they
found that
$\Delta \Sigma = (100\pm30)h M_\odot$~pc$^{-2}$ on scales of $94h^{-1}$~kpc.  With this constraint the 
adiabatically compressed models are again strongly favored (by 1000 to 1).  In general, any third
constraint that is dominated by the contribution from one mass component will break the degeneracy and
lead to constraints on the degree of adiabatic compression or an additional structural variable
such as the inner slope of the dark matter density distribution.

The sample of lenses available for such analyses will continue to grow and can include lenses with time delay 
measurements (which constrain the halo structure by measuring the surface density near the lensed images, 
Kochanek 2002) as well as those with velocity dispersions.  With larger samples it should be possible to explore 
additional correlations such as the scaling of the stellar mass fraction and mass-to-light ratios with halo mass 
and the dependence
of the evolution rate on halo mass.  In the Mandelbaum et al. (2006) and Padmanabhan (2004) analyses of early-type
galaxies in the SDSS, the changes in the mass-to-light ratio with halo mass are due to an increasing dark matter
fraction with mass rather than changes in the stellar populations, but their results depend on population 
synthesis models to correctly estimate the stellar masses.  In a larger sample of lenses, this could be tested
directly.  \citet{md06} and \citet{tkbbm06} see some evidence for differential evolution with mass, but significantly
larger samples will be needed to test this given the sensitivity of even the present results to sample weighting.

\acknowledgements

We would like to that L. Koopmans, L. Moustakas, E. Rozo and T. Treu for their comments
and R. Mandelbaum for discussions of weak lensing.

\clearpage




\begin{deluxetable}{lccccccclcccc}
\rotate
\tablewidth{\vsize}
\tablecaption{Lens data}
\tabletypesize{\scriptsize}
\startdata
\hline
\hline
$\rm {Objects}$ & $z_{\rm l}$ & $z_{\rm s}$ & ${\rm R}_{\rm e}$ & ${\rm R}_{\rm E}$ & ${\rm M}_{\rm E}$ & $\sigma_{\rm ap}$
& ${\rm L}_{\rm B}$ & $\rm Aperture$  & $\rm seeing$ & $f_{dm}(<R_E)$ & $f_{dm}(<R_E)$ &$\rm {reference}$ \\
& & & $('')$ & $(\rm {kpc})$ & $(10^{10})$\msun & $(\rm {km\,s^{-1}})$ & $(10^{11})$\lsun & $('')$ & $('')$ & ${\rm compr}$ &
${\rm no compr}$ & \\
\hline

SDSS J0037$-$0942 & 0.1955 & 0.6322 & 2.38 & 4.77  & 27.3  & 265$\pm$10(30) & 1.24$\pm$0.11 & 1.5  & 1.76 & 0.42 & 0.22 &1,2\\
SDSS J0216$-$0813 & 0.3317 & 0.5235 & 3.37 & 5.49  & 48.2  & 332$\pm$23(37) & 2.60$\pm$0.39  & 1.5  & 2.73& 0.52 & 0.35 &1,2\\
SDSS J0737+3216   & 0.3223 & 0.5812 & 3.26 & 4.83  & 31.2  & 310$\pm$15(30) & 2.13$\pm$0.16 & 1.5  & 2.33 & 0.42 & 0.25 &1,2\\
SDSS J0912+0029   & 0.1642 & 0.3240 & 4.81 & 4.55  & 39.6  & 313$\pm$12(29) & 1.63$\pm$0.20 & 1.5  & 1.49 & 0.50 & 0.32 &1,2\\
SDSS J0956+5100   & 0.2405 & 0.4700 & 2.60 & 5.02  & 37.0  & 299$\pm$16(30) & 1.25$\pm$0.18 & 1.5  & 1.63 & 0.43 & 0.24 &1,2\\
SDSS J0959+0410   & 0.1260 & 0.5349 & 1.82 & 2.25  & 7.7   & 212$\pm$12(21) & 0.27$\pm$0.06 & 1.5  & 1.58 & 0.39 & 0.20 &1,2\\
SDSS J1250+0523   & 0.2318 & 0.7950 & 1.77 & 4.26  & 18.9  & 254$\pm$14(25) & 1.13$\pm$0.06 & 1.5  & 1.48 & 0.36 & 0.16 &1,2\\
SDSS J1330$-$0148 & 0.0808 & 0.7115 & 1.23 & 1.30  & 3.2   & 178$\pm$9(17)  & 0.09$\pm$0.06 & 1.5  & 1.84 & 0.29 & 0.10 &1,2\\
SDSS J1402+6321   & 0.2046 & 0.4814 & 3.14 & 4.66  & 30.3  & 275$\pm$15(28) & 1.06$\pm$0.23 & 1.5  & 1.83 & 0.44 & 0.25 &1,2\\
SDSS J1420+6019   & 0.0629 & 0.5352 & 2.60 & 1.27  & 3.9   & 194$\pm$5(17)  & 0.28$\pm$0.09 & 1.5  & 2.37 & 0.33 & 0.13 &1,2\\
SDSS J1627$-$0053 & 0.2076 & 0.5241 & 2.14 & 4.11  & 22.2  & 275$\pm$12(26) & 0.75$\pm$0.07 & 1.5  & 1.87 & 0.34 & 0.15 &1,2\\
SDSS J1630+4520   & 0.2479 & 0.7933 & 2.02 & 7.03  & 50.8  & 260$\pm$16(29) & 1.15$\pm$0.33 & 1.5  & 1.52 & 0.57 & 0.42 &1,2\\
SDSS J2300+0022   & 0.2285 & 0.4635 & 1.80 & 4.56  & 30.4  & 283$\pm$18(30) & 0.72$\pm$0.13 & 1.5  & 1.87 & 0.40 & 0.20 &1,2\\
SDSS J2303+1422   & 0.1553 & 0.5170 & 4.20 & 4.41  & 27.5  & 260$\pm$15(26) & 1.09$\pm$0.22 & 1.5  & 1.65 & 0.50 & 0.32 &1,2\\
SDSS J2321$-$0939 & 0.0819 & 0.5324 & 4.47 & 2.43  & 11.7  & 236$\pm$17(21) & 0.76$\pm$0.12 & 1.5  & 2.09 & 0.44 & 0.25 &1,2\\
0047$-$281        & 0.484  & 3.595  & 0.82 & 12.43 & 58.0  & 250$\pm$30(41) & 1.20$\pm$0.08 & 1.5$\times$4.3& 0.7& 0.59 &0.44 &3,4\\
C0302+006         & 0.938  & 2.941  & 1.6  & 10.6  & 67.0  & 256$\pm$19(31) & 3.25$\pm$0.45 & 0.5$\times$1.25&0.8& 0.29 &0.11 &8,9\\
PG1115+080        & 0.310  & 1.722  & 0.85 & 4.74  & 17.0  & 281$\pm$25(36) & 0.35$\pm$0.02 & 1.0$\times$1.0   & 0.8 &0.83 &0.78 &5,6\\
H1417+526         & 0.810  & 3.399  & 1.06 & 11.4  & 70.8  & 212$\pm$18(26) & 2.42$\pm$0.23 & 0.32$\times$1.25 & 0.75&0.62 &0.49 &9,10 \\
H1543+535         & 0.497  & 2.092  & 0.41 & 2.4  & 3.4  & 108$\pm$14(17) & 0.28$\pm$0.04 & 0.3$\times$1.25 & 0.8 &0.84  &0.78 &9\\
MG1549+305        & 0.111  & 1.170  & 0.82 & 2.33  & 12.0  & 227$\pm$18(28) & 0.17$\pm$0.02 & 1.0$\times$4.3  & 0.65 &0.60 &0.46 & 11,12\\
MG2016+112        & 1.004  & 3.263  & 0.31 & 13.70 & 110.0 & 304$\pm$27(47) & 1.60$\pm$0.08 & 0.65             & 0.7 &0.44 &0.26 & 7\\
\enddata
\tablecomments{\label{tab:tab1} 
$z_{\rm l}$ 
and $z_{\rm s}$ are the lens and source redshifts, ${\rm R}_{\rm e}$ is the lens effective radius, ${\rm R}_{\rm E}$ and ${\rm M}_{\rm E}$
are the lens Einstein radius and Einstein mass,  $\sigma_{\rm ap}$ is the measured velocity dispersion. We include both the measurement
 errors inside the listed aperture
and our estimated systematic uncertainties are in (brackets).
We lacked the seeing FWHM  for SDSS J1627$-$0053 and SDSS J2300+0022 and simply used the average value for the other SDSS objects.
$f_{dm}(<R_E)$ is the projected dark matter fraction inside the Einstein radius for adiabatically compressed (``compr") or not compressed 
(``no compr") models.\\
\citet{tkbbm06}(1), \citet{ktbbm06}(2),  \citet{tksse03}(3), \citet{kt03}(4) \\ \citet{tonry98}(5), \citet{tk02a}(6)
, \citet{kt02}(7) \\ \citet{tk03}(8), \citet{tk04}(9) , \citet{kkf98}(10), \citet{llslb93}(11) \\ \citet{lclsl96}(12) }
\end{deluxetable}
\clearpage

\begin{figure}
\epsscale{.80}
\plotone{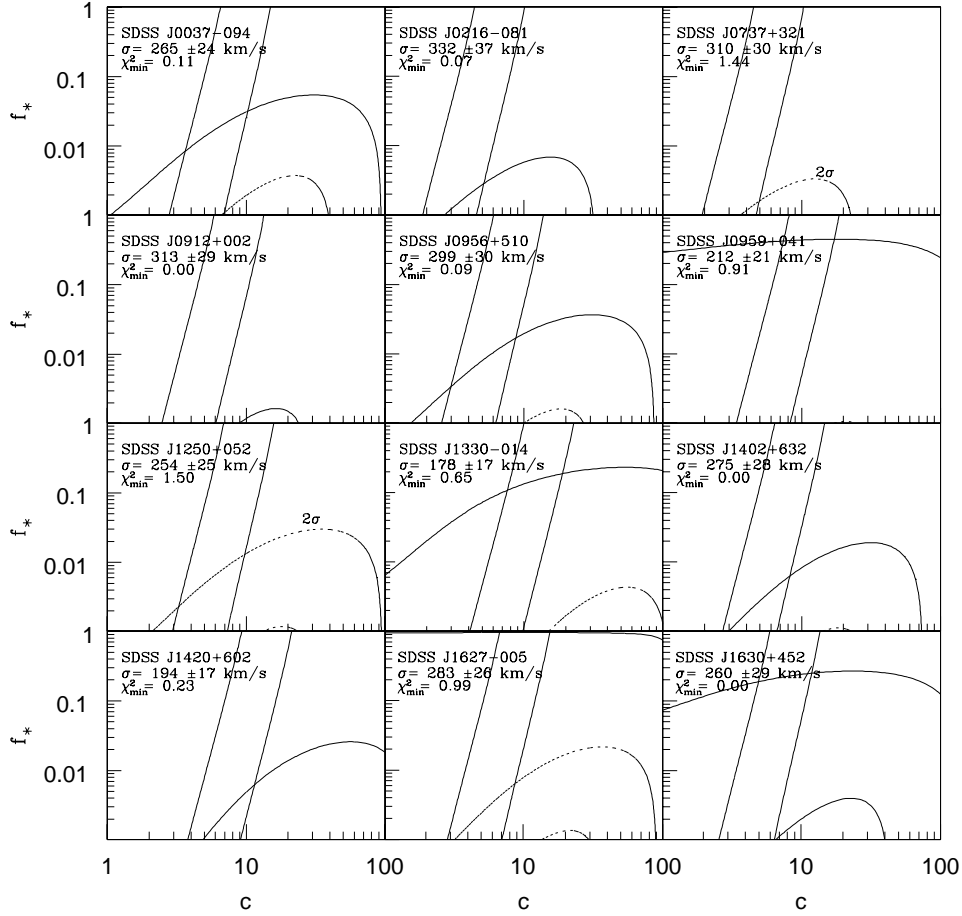}
\caption{\label{fig:figure1}
The goodness of fit to
the velocity dispersion as a function of stellar mass fraction
$f_*$ and concentration $c$. The solid curves are drawn at $\Delta\chi^2 = 1 (1\sigma)$
for the fit to the velocity dispersion and the dotted lines are drawn at $\Delta\chi^2 = 4$, 9 and 16 (2, 3, and 4 $\sigma$). When there is no region with
$\Delta\chi^2 < 1$, we label the lowest contour present.
The roughly vertical pair of solid lines indicate the $1\sigma$ range of concentrations given the halo mass at each
point. The inset text identifies the object, the measured velocity dispersion and the $\chi^2$ of the best fit. These are the isotropic ($\beta=0$)
 adiabatically compressed models that include our estimate of the systematic uncertainties in the stellar dynamical measurements.
}
\end{figure}
\clearpage

\addtocounter{figure}{-1}

\begin{figure}
\epsscale{.80}
\plotone{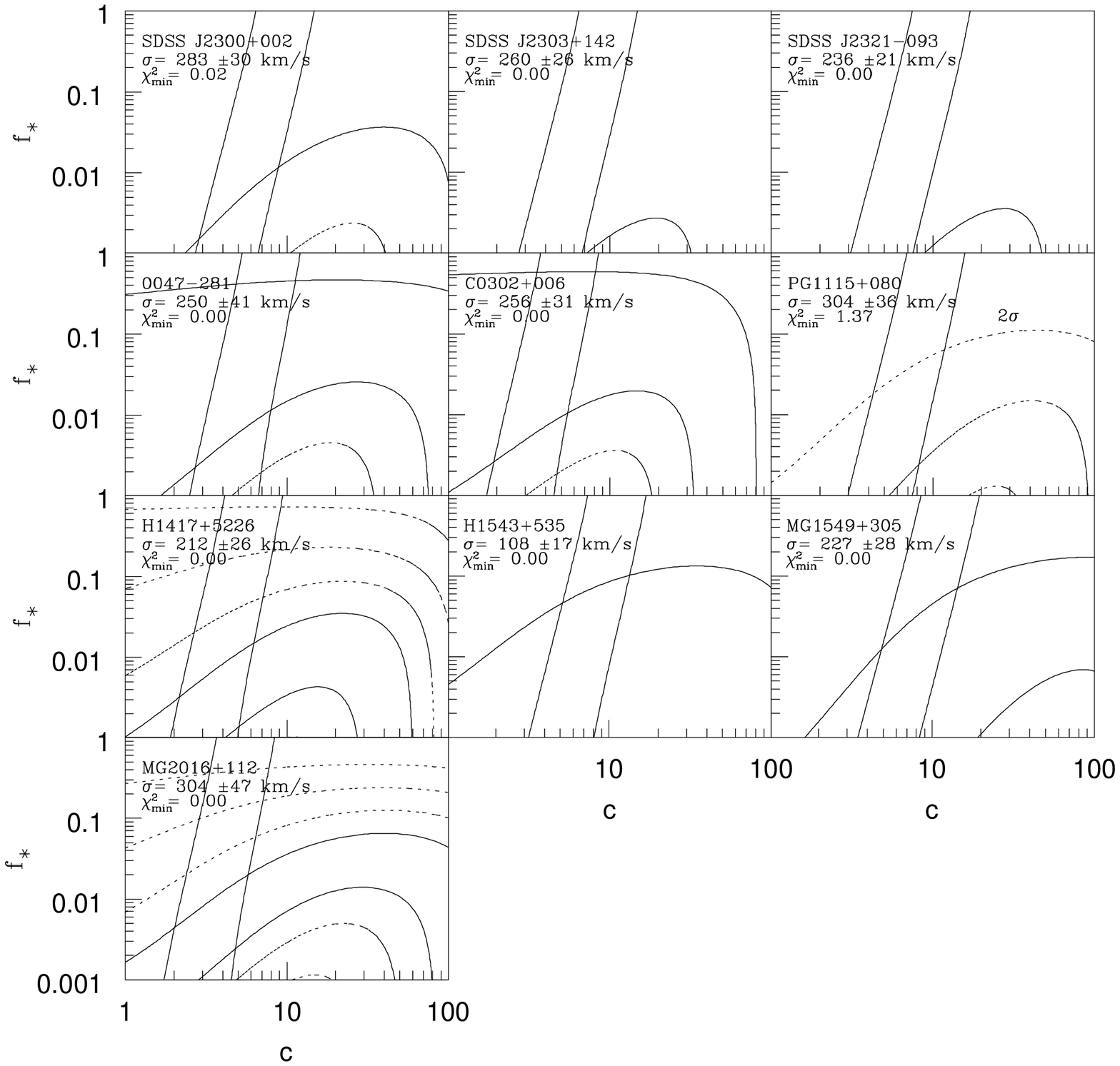}
\caption{continued
}
\end{figure}
\clearpage

\begin{figure}
\epsscale{.80}
\plotone{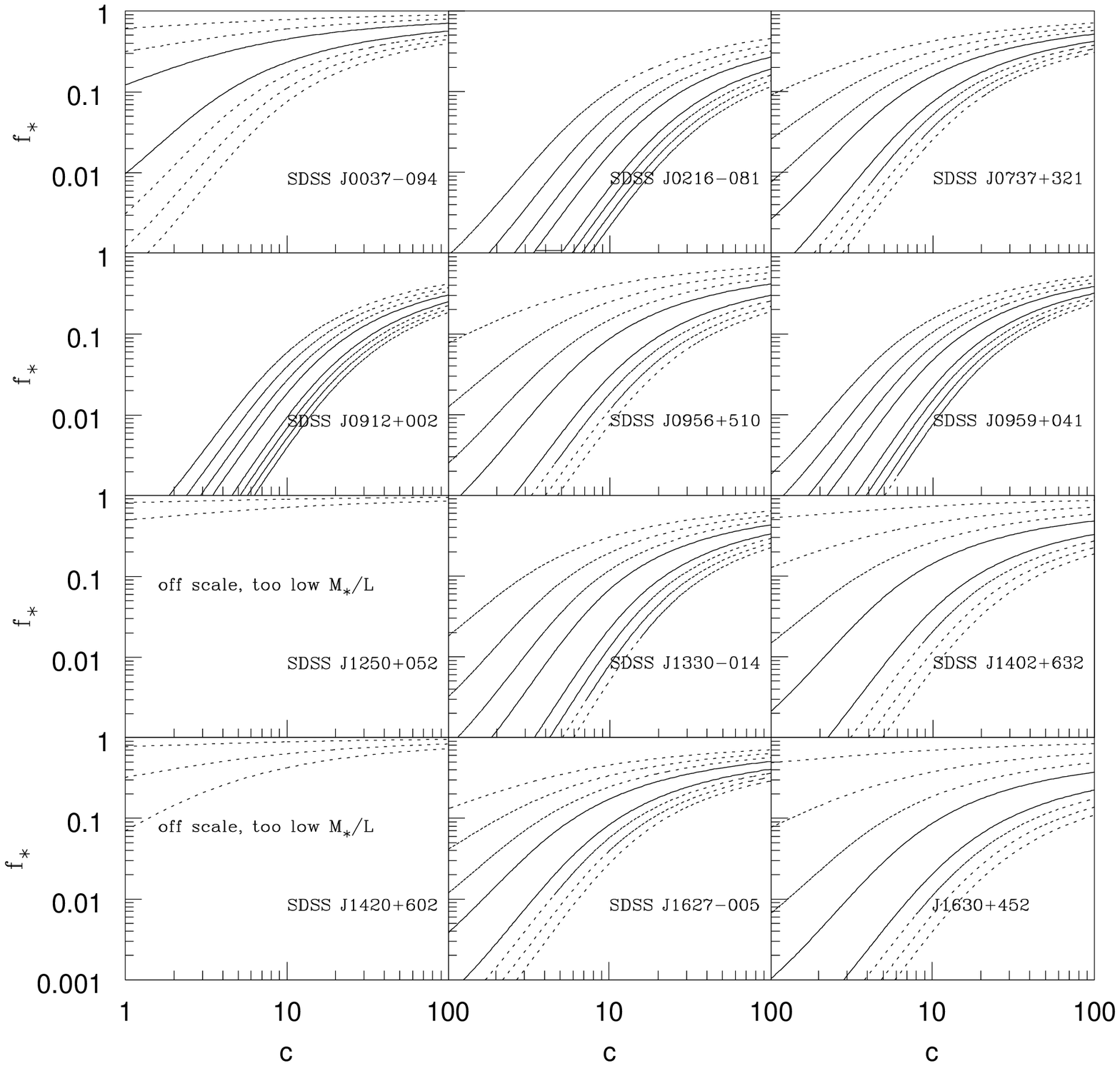}
\caption{\label{fig:figure3}
The goodness of fit to the mean trend in the stellar mass-to-light ratio as a function of the stellar mass fraction and the concentration $c$.
The solid curves are drawn at $\Delta\chi^2 = 1$ $(1\sigma)$
for the fit to the mass-to-light ratio and the dotted lines are drawn at $\Delta\chi^2 = 4$, 9 and 16 (2, 3, and 4 $\sigma$).
The inset text identifies the object. These are the isotropic ($\beta=0$)
 adiabatically compressed models that include our estimates of the systematic uncertainties in the stellar dynamical measurements. 
For these figures, we use the best fit evolution
model -- the fits would improve if we included the measurement errors in the evolution model.
}
\end{figure}
\clearpage

\addtocounter{figure}{-1}

\begin{figure}
\epsscale{.80}
\plotone{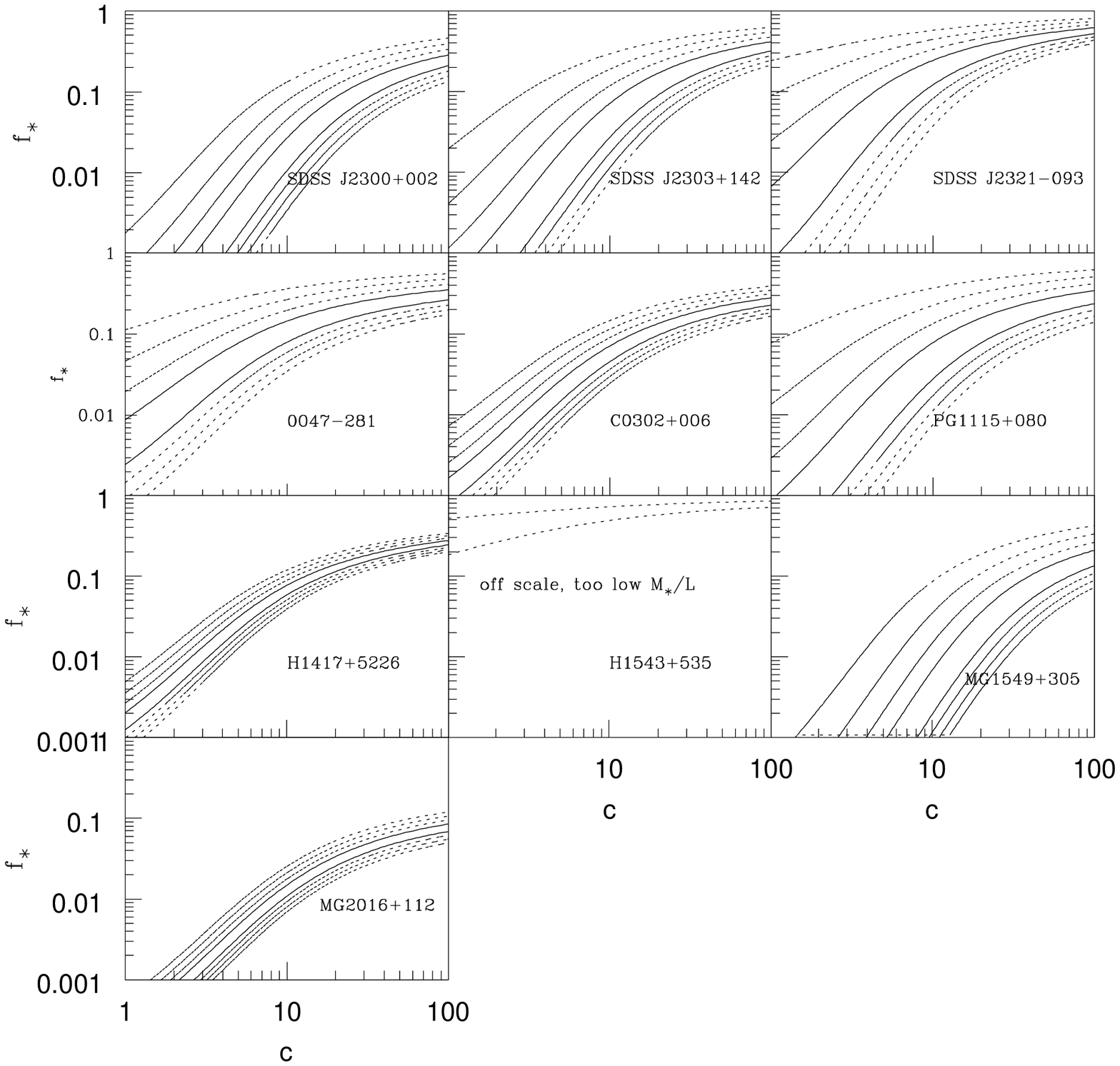}
\caption{continued
}
\end{figure}
\clearpage

\begin{figure}
\epsscale{.80}
\plotone{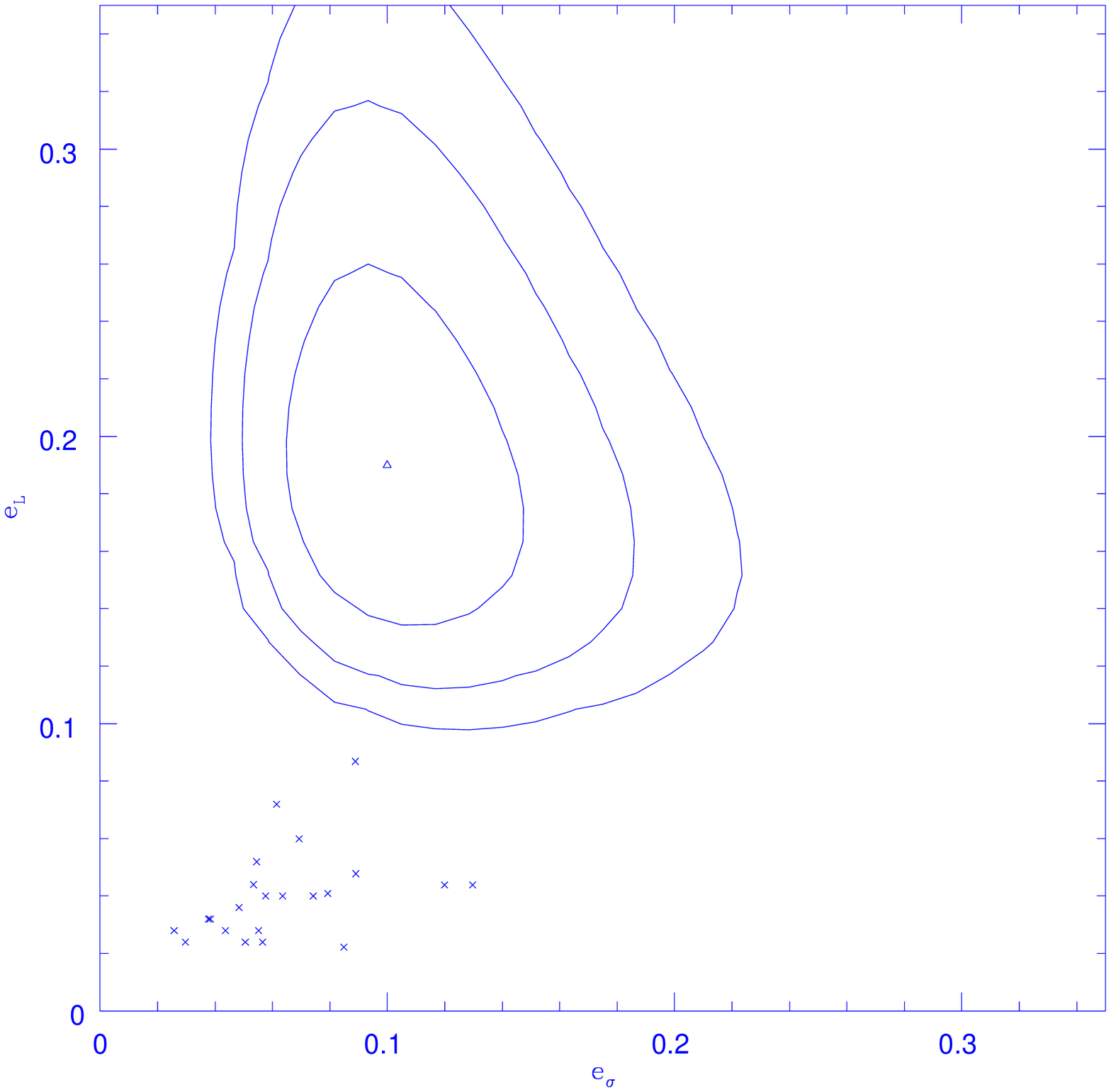}
\caption{\label{fig:figure6}
The probability distribution of the fractional systematic errors $e_\sigma$ and $e_L$ in the velocity
dispersion and mass-to-light ratio.  The contours encompass $68\%$, $95\%$ and
$99.7\%$ of the probability starting from the maximimum likelihood solution indicated
by the triangle.  The crosses indicate the measurement errors from
from \citet{tkbbm06} and \citet{ktbbm06}.
}
\end{figure}
\clearpage

\begin{figure}
\epsscale{.80}
\plotone{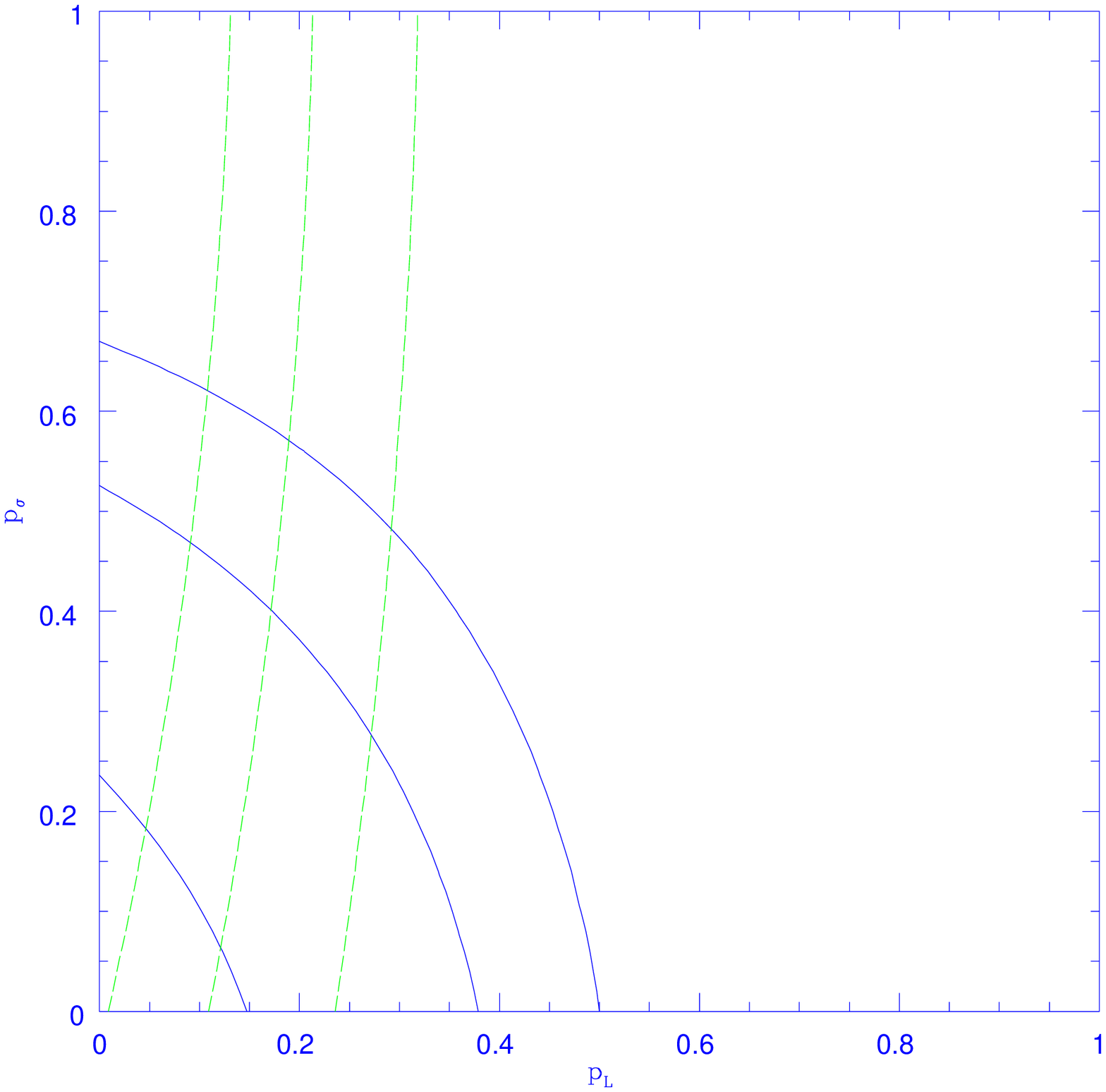}
\caption{\label{fig:figure5}
The likelihood distributions for the probability that the galaxy sample is homogeneous in its dynamics ($p_\sigma$)
or its evolution ($p_L$).  The contours encompass 68\%, 95\% and 99.7\% of the probability.
The solid contours use
the measurement errors for the dynamical uncertainties while the dashed contours include our estimate
of the systematic uncertainties in the dynamical measurements.  In both cases we used an adiabatically
compressed, isotropic ($\beta=0$) model.
}
\end{figure}
\clearpage

\begin{figure}
\epsscale{.80}
\plotone{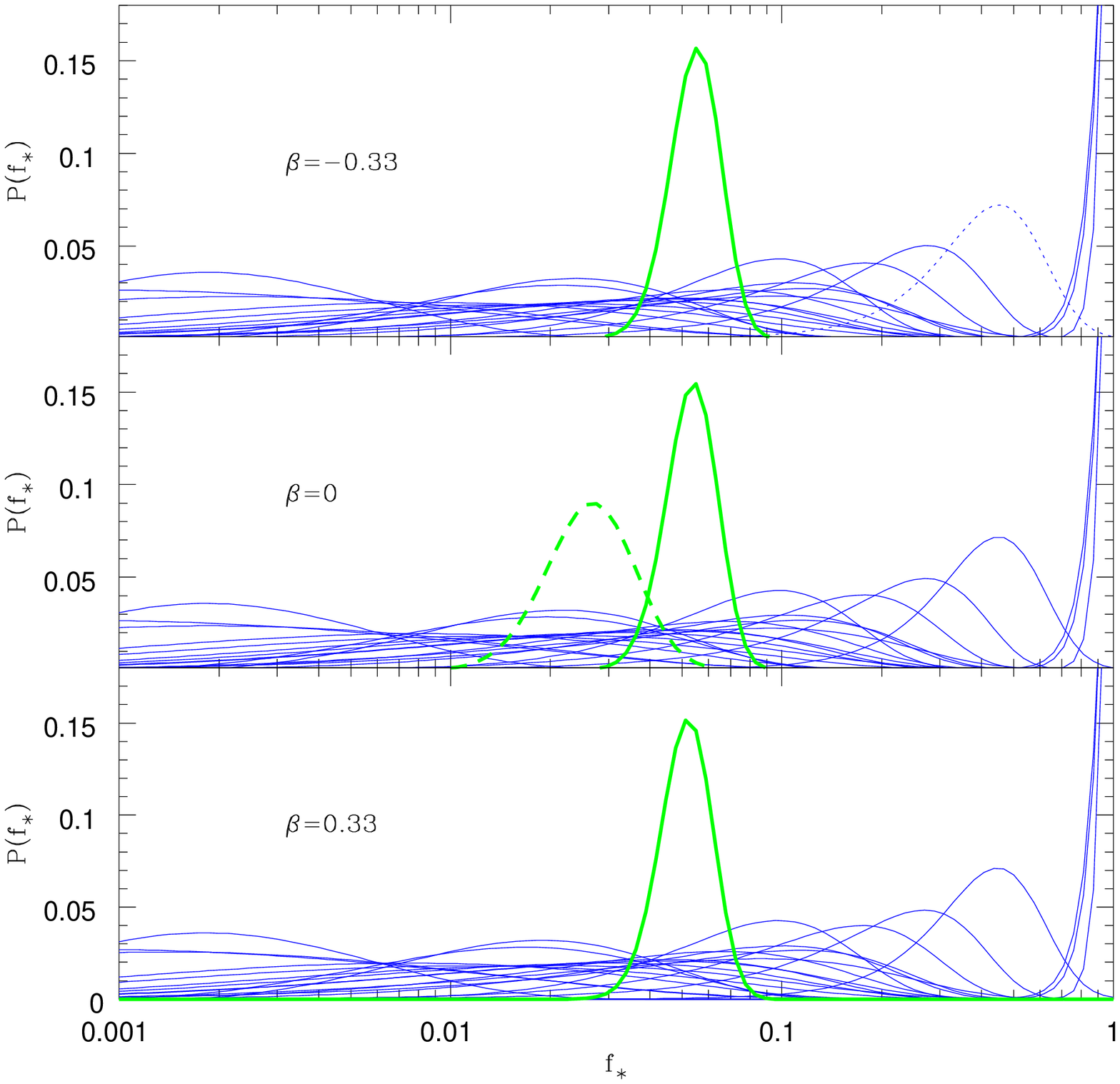}
\caption{\label{fig:figure7}
The probability distribution for the stellar mass fraction $f_*$ 
 for tangentially anisotropic ($\beta=-0.33$, top panel), isotropic ($\beta=0$, middle), 
and radially anisotropic ($\beta=0.33$, bottom) dynamical models.
The thin lines in each panel show the weak constraints found for the individual galaxies,
and the thick solid line corresponds
to the joint probability from combining the sample. These models are adiabatically compressed using the modified errors and include
the fit to the mass-to-light ratios. 
The thick dashed line in the middle panel shows the effect of not including the adiabatic compressions.
}
\end{figure}
\clearpage

\begin{figure}
\epsscale{.80}
\plotone{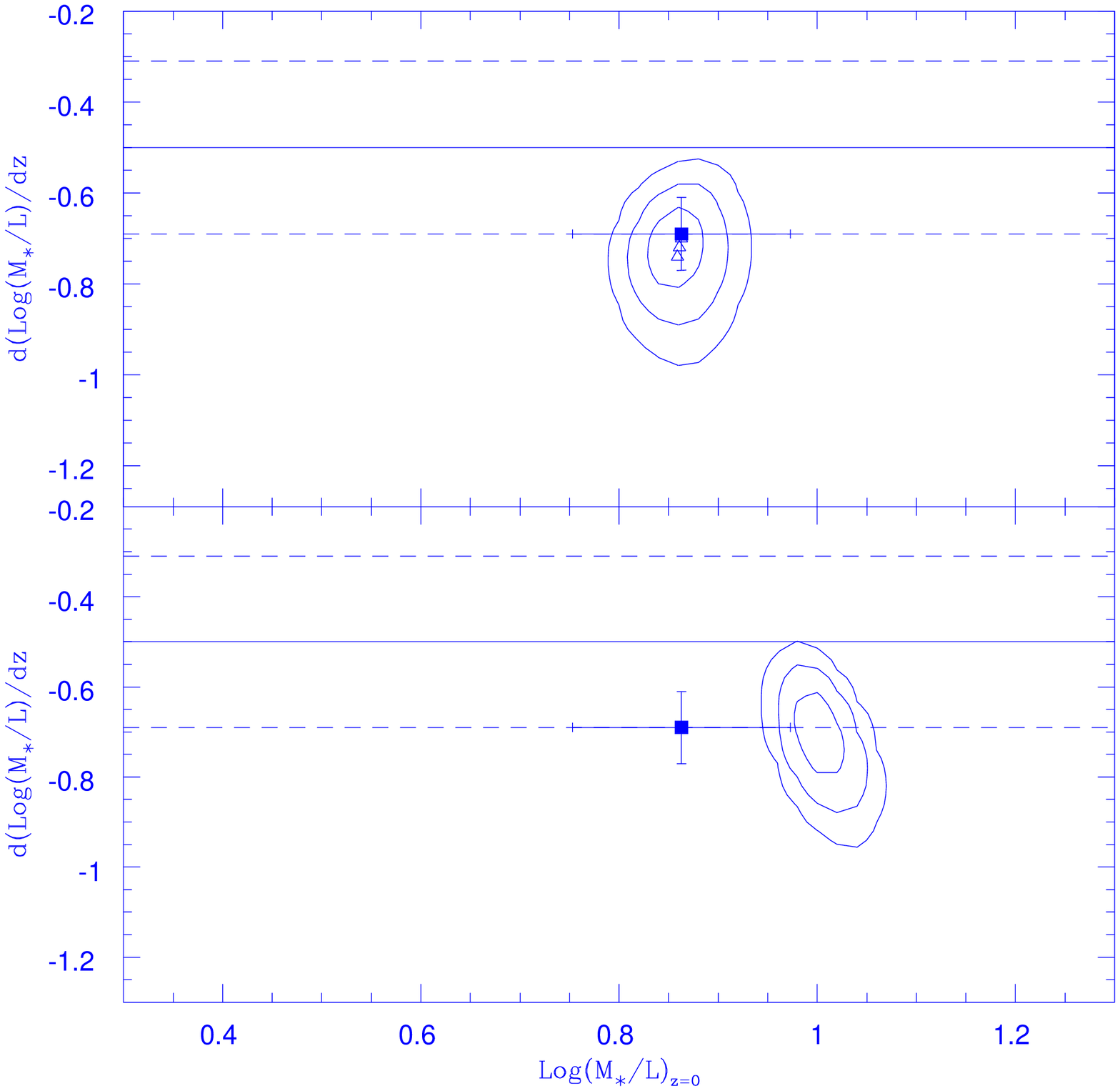}
\caption{
\label{fig:figure8}
The probability distributions for the local mass-to-light ratio $(M_{\ast}/L)_0$ and its evolution 
$d\log(M_{\ast}/L)/dz$ in the adiabatically compressed (top) and uncompressed (bottom) models. The contours show the $68\%$, $95\%$, and $99.7\%$
enclosed probability contours for the isotropic models.  The estimated evolution rate is marginally inconsistent with the
estimated of $d\log(M_{\ast}/L)/dz=-0.50\pm0.19$ from \citet{rk05} which is shown by the horizontal band
of solid and dashed lines.  The three triangles in each panel show the effect of changing the isotropy on 
the likelihood peak, with $\beta=-0.33$, $\beta=0$, and $\beta=0.33$  as we move from upper left to lower right.
The squares with error bars are the results from  \citet{tkbbm06} for the same galaxies.
}
\end{figure}
\clearpage

\begin{figure}
\epsscale{.80}
\plotone{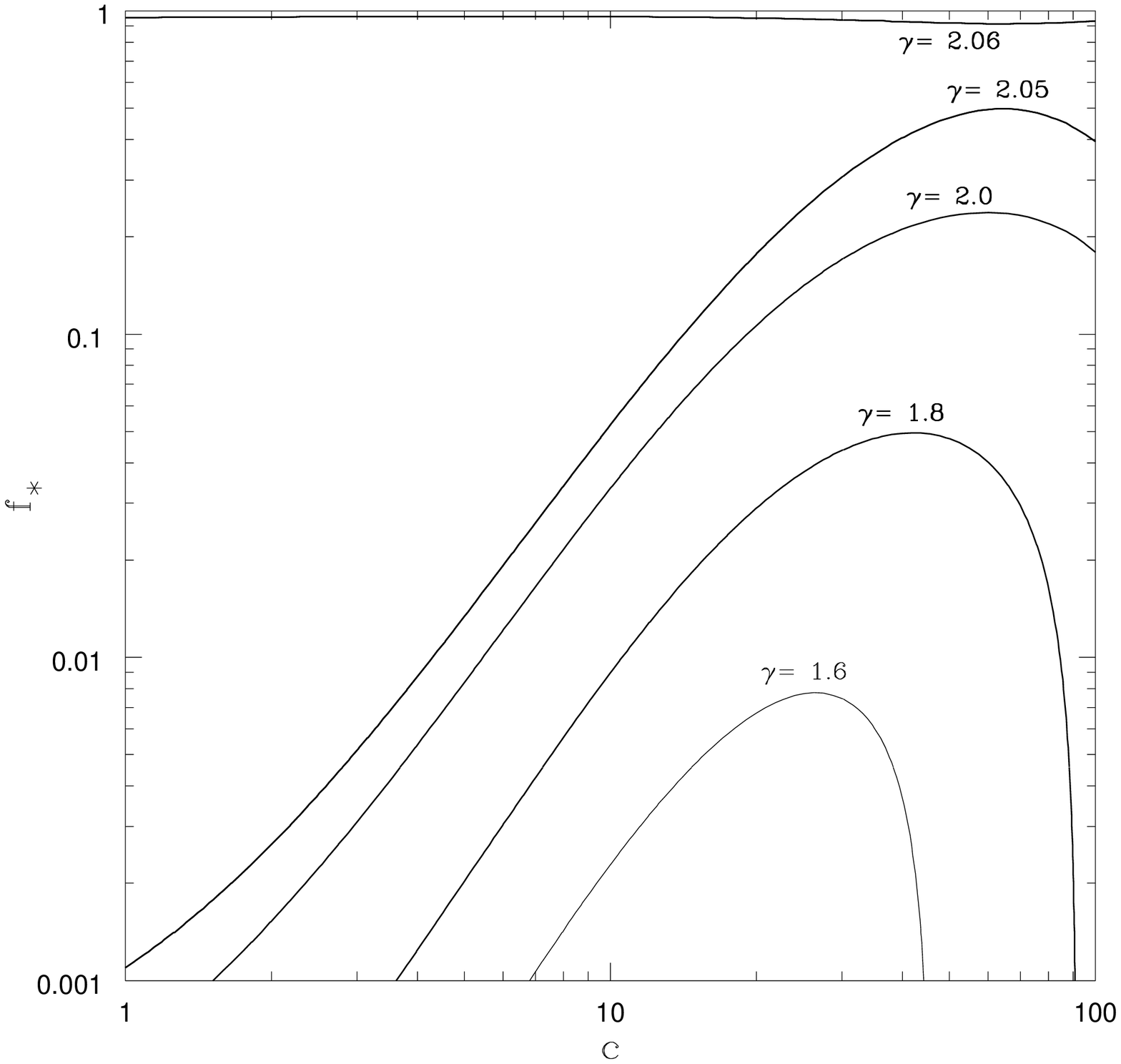}
\caption{
\label{fig:figure9}
The range for the density slope exponent $\gamma$, where $\rho\varpropto r^{-\gamma}$, for the typical lens SDSS J0037--0942. 
We estimated $\gamma$ by fitting the projected mass distribution as a power law between $R_e/8$ and $R_E$. Note that the variation of 
$\gamma$ over the  physically interesting regime is comparable to the scatter observed by \citet{ktbbm06} of 
$1.8\la \gamma \la 2.3$.}
\end{figure}

\end{document}